\documentclass[%
 prl,twocolumn,
 nofootinbib,
groupedaddress,
aps,
]{revtex4-1}

\usepackage{amsmath, amssymb}
\usepackage{graphicx,xcolor}% Include figure files
\usepackage{dcolumn}% Align table columns on decimal point
\usepackage{bm}% bold math
\usepackage{hyperref}% add hypertext capabilities
\newcommand{\p}{\hat{\mathbf{p}}}
\newcommand{\n}{\hat{\mathbf{n}}}

\begin{document}

%\preprint{APS/123-QED}

\title{Phase decoherence of gravitational wave backgrounds}

\author{Aoibheann Margalit}
\email{a.margalit19@imperial.ac.uk}
\author{Carlo~R.~Contaldi}%
\author{Mauro Pieroni}%
\affiliation{%
 Blackett Laboratory, Imperial College London, London, SW7 2AZ, UK}
 

\date{\today}

\begin{abstract}
Metric perturbations affect the phase of gravitational waves as they propagate through the inhomogeneous universe. 
This effect causes Stochastic Gravitational Wave Backgrounds (SGWBs) to lose {\sl any} phase coherence that may have been present at emission or horizon entry. 
We show that, for a standard cosmological model, this implies complete loss of coherence above frequencies $f \sim 10^{-12}$ Hz. 
The result is that any attempts to map SGWBs using phase-coherent methods have no foreseeable applications. 
Incoherent methods that solve directly for the {\sl intensity} of the SGWBs are the only methods that can reconstruct the angular dependence of any SGWB.
\end{abstract}

\pacs{Valid PACS appear here}% PACS, the Physics and Astronomy
\maketitle

{\sl Introduction.--} Gravitational Wave (GW) observatories such as LIGO \cite{ligo2015, ligo2016a, ligo2016b} and LISA \cite{lisa2011, lisa2012, lisa2016} constitute coherent detectors of the underlying observable, the polarised metric strain $h$. 
In principle, they are able to measure both amplitude and phase of the time-evolving strain field as waves pass through the detector. 
This is analogous to radio frequency antennas that measure the amplitude and phase of a passing electromagnetic wave. 
In fact, the ability of a dipole antenna to measure the phase of an electromagnetic wave underpins the method of encoding signals in Frequency Modulated (FM) radio transmissions which are notoriously more robust to interference than simple Amplitude Modulated (AM) radio transmissions. 
Since GW observatories act like FM radio receivers it is useful to consider what information is carried by the phase of the signal.

The phase of the GW signal arriving from any particular direction on the sky is very useful in the case of point sources such as the signal arising from mergers of massive, compact, astrophysical objects. 
In this case the phase information  can be used to differentiate the signal from the background noise as long as the phase response of the detector is understood well enough. The benefit is a much improved ability to constrain the directional dependence of the signal above the raw angular resolution of the detector. This is particularly the case when using networks of multiple detectors \cite{ballmer2005}. Methods where the phase information is being used to solve for the signal are classed as phase-coherent methods \cite{cornish2014, gair2014, romano2015}. These solve directly for the amplitude of the polarised strain such as $h_+$ and $h_\times$ by considering integration of the detector output that is {\sl linear} in the detector response.

Phase-coherent methods have been considered also in attempts to map Stochastic Gravitational Wave Backgrounds (SGWBs) \cite{cornish2014, gair2014, romano2015}. 
SGWBs can be generally placed into two classes. 
SGWBs of the first class are generated by the overlapping signals from many individual, usually astrophysical, sources that lie below some effective confusion limit of the detector \cite{regimbau2011, zhu2011, rosado2011, marassi2011}. 
The confusion limit for any particular detector is determined by either time-domain or angular resolution in the detector response. 
This class of SGWBs is not expected to be phase-coherent --- the phases of any signals arriving from any two directions on the sky should not be correlated \cite{Cusin:2017mjm, Conneely:2018wis}. 
At the very least, for signals generated by sub-horizon mechanisms (including all astrophysical events and some cosmological sources like phase transitions), causality should ensure there is no phase-coherence over large angular scales. 

The second class are primordial backgrounds. 
This is the case where the SGWB was seeded by a mechanism which is coherent on cosmological scales  \cite{binetruy2012, caprini2018}.
The important difference is that, in this case, waves at each wavelength (frequency) start oscillating (and traveling from cosmological distances to today) in phase irrespective of the direction in which they propagate. 
In principle such a situation will lead to a very different SGWB than the previous case and it was recognised very early on that a phase-coherent inflationary SGWB would form {\sl standing waves} \cite{grishchuk1975}. 
The standing wave nature of a SGWB is a feature that could be directly tested by coherent detectors \cite{Contaldi:2018akz}. 
This is directly analogous to the angular coherence of Cosmic Microwave Background (CMB) anisotropies which results in the acoustic peak signal \cite{boomerang2006,wmap2012,planck2019}. The important distinction is that CMB photons carry no temporal phase coherence, meaning that any phase information collected by radio interferometers is routinely discarded in the CMB mapping process. Meanwhile, it was thought that any coherency in the phase information collected by GW interferometers would be a smoking gun for a primordial background, as described above.

In this {\sl article} we argue that there is no foreseeable scenario where phase-coherent estimators are useful for mapping SGWBs. 
It is well known that GWs will be affected by perturbations in the background metric along which they travel. 
At observable frequencies the geometric limit \cite{isaacson1968a, isaacson1968b} can be used to calculate the effect of perturbations at linear order \cite{laguna2010,contaldi2017,bartolo2019a,pitrou2019}. 
The result is that perturbations along the line-of-sight (l.o.s.) cause decoherence \cite{Bartolo:2018evs,bartolo2019b,dimastrogiovanni2020} of any phase information in any SGWB. 
One could argue that the phase shifts induced by the perturbations carry useful information themselves, however, as we will show in a detailed calculation of the effect, the decoherence is so strong that it randomises the phase information very efficiently at observable frequencies. 
The consequence is that only {\sl phase-incoherent} methods for reconstructing any SGWB map are of any use\footnote{While this fact had already been recognised for astrophysical SGWBs \cite{Cusin:2017mjm, Conneely:2018wis}, we emphasise that our results show that this applies also to primordial or cosmological SGWBs.}.
These are methods that use the square of the detector response to solve directly for the {\sl intensity} of the underlying strain field \cite{renzini2018,renzini2019}. 
These methods assume no phase coherence in the data and are analogous to radio frequency methods for mapping CMB anisotropies using coherent radio detectors \cite{Myers:2002tn}.

{\sl Line-of-sight decoherence.--}
We follow Isaacson's geometric optics approach \cite{isaacson1968a, isaacson1968b} and decompose the metric as 
$g_{\mu \nu} = \gamma_{\mu \nu} + \epsilon h_{\mu \nu}$\,,
where, in our case, $\gamma_{\mu \nu}$ is a flat 
Friedmann–Lema\^itre–Robertson–Walker (FLRW) metric ($c = 1$) with scalar perturbations $\Phi(\eta, \mathbf{x})$ and  $\Psi(\eta, \mathbf{x})$,
\begin{equation}\label{eq: background metric}
    \gamma_{\mu \nu} \textrm{d}x^{\mu} \textrm{d}x^{\nu}\! = a^2(\eta)\!\left[ - ( 1 + 2 \Phi) \textrm{d}\eta^2 + ( 1 - 2 \Psi) \textrm{d} \mathbf{x}^2\right].
\end{equation}
Here, $h_{\mu \nu}$ are GW perturbations on top of this background and $\epsilon$ is a small expansion parameter. 
In this limit, we assume that the GW wavelength is much smaller than the curvature scale set by the total metric.
We neglect the back-reaction of the GWs on the background spacetime by setting its stress tensor to zero.
We define $\bar{h}_{\mu \nu} = h_{\mu \nu} - \frac 12 \gamma_{\mu \nu} \gamma^{\rho \sigma} h_{\rho \sigma}$ and, choosing the transverse-traceless gauge, we write $\bar{h}^{\lambda}_{\mu \nu} = \mathcal{A} e^{\lambda}_{\mu \nu} e^{i \varphi/ \epsilon}$.
Here, $\lambda \in \{+, \times\}$ labels the polarisation described by the tensor $e_{\mu \nu}^{\lambda}$ while $\mathcal{A}$ and $\varphi$ are real functions of retarded time corresponding to the amplitude and phase of the GW. 
The GW wavevector can then be identified as $k_{\mu} = \partial_{\mu} \varphi$.

\begin{figure}[t]
    \centering
    \includegraphics[width = \columnwidth]{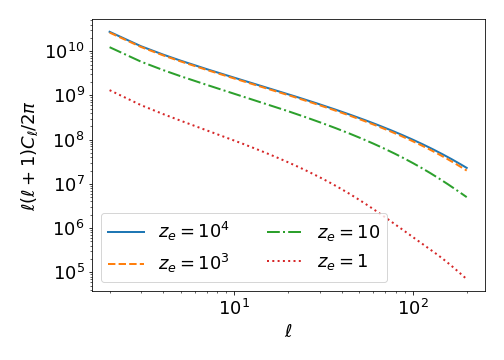}
    \caption{Time delay power spectrum $C_\ell^{\delta \varphi}$ for single-redshift GW sources and an observing frequency $f=10^{-9}$ Hz. The power peaks on the largest angular scales since the effect is a direct integration of the underlying potential.}
    \label{fig: plot of C_l}
\end{figure}

With this notation, Einstein's equations provide two constraints on the wavevector \cite{isaacson1968a, laguna2010}.
At the order of $\epsilon^{-2}$, we find $k^2 = 0$, i.e. the GW follows a null geodesic. 
Encoded in this is the fact that the phase $\varphi$ is constant during propagation since, for affine parameter $l$, $d\varphi/dl = k^{\mu} \nabla_{\mu} \varphi = k^{\mu}k_{\mu} = 0$.
At $\mathcal{O}(\epsilon^{-1})$, we find $k^{\mu} \nabla_{\mu} e^{\lambda}_{\rho \sigma} = 0$, i.e. the polarisation tensor is parallel-transported along the null geodesic.
To leading order in the scalar perturbations we have $k_{\mu} = 2 \pi \, f (1, \p)$ where $f$ is the intrinsic frequency of the GW and $\p$ is the unit vector in its direction of travel.
Note that the angular frequency measured by a comoving observer with 4-velocity $u^{\mu} =  (1, \mathbf{0})/a$ is $\omega(a) = k_{\mu} u^{\mu} = 2 \pi \, f/a$.
That means, from our position at scale factor $a=1$, $f$ is the same as the GW frequency measured in our detectors.

To account for scalar perturbations to the metric at linear order, one must solve the geodesic equation for the vector $k_{\mu}$ in the background $\gamma_{\mu \nu}$. 
The resulting phase shift accumulated along the l.o.s. is \cite{laguna2010}
\begin{equation}\label{eq: phase shift}
    \delta \varphi(\n) = 2\pi \, f \int\limits_{\text{l.o.s.}} \left[\Phi(\eta, \mathbf{x}) + \Psi(\eta, \mathbf{x})\right]\, \textrm{d} \eta\,,
\end{equation}
where the integral runs from conformal time at emission $\eta_e$ to observation $\eta_o$ and follows the null trajectory of the GW in a universe without perturbations, $\mathbf{x} = (\eta_o - \eta) \n$.
Here, $\n = -\p$ denotes the direction on the sky.
This has a natural interpretation in terms of the cosmological Shapiro time delay.
During propagation, the GW is deflected by gravitational wells along its path causing it to travel an extra distance $d(\n) = \int (\Phi + \Psi) \textrm{d} \eta$. 
The phase shift is due to the additional non-integer number of cycles the wave experiences along this detour compared to the unperturbed path.
In particular, we see that $\varphi$ is no longer conserved along the geodesic.

We quantify this effect for a standard cosmological model and discuss its implications for the detection of SGWBs. As we will show, the measured phase $\varphi_o$ (in units of $2 \pi$) is randomised to such an extent that any information contained in the initial phase distribution is scrambled for all observable frequencies.

\begin{figure}[t]
    \centering
    \includegraphics[width = \columnwidth]{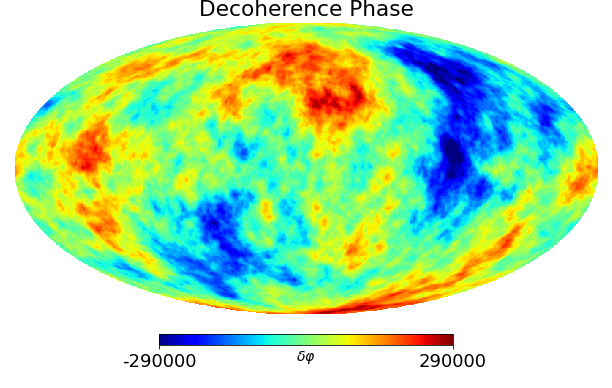}
    \caption{A realisation of $\delta\varphi$ for $z_e=1000$ at an observing frequency $f=10^{-9}$ Hz. The phase shift along any l.o.s. is typically orders of magnitude larger than a single cycle.}
    \label{fig:dvarphi}
\end{figure}

\begin{figure*}[t]
\centering
    \includegraphics[width=0.25\linewidth]{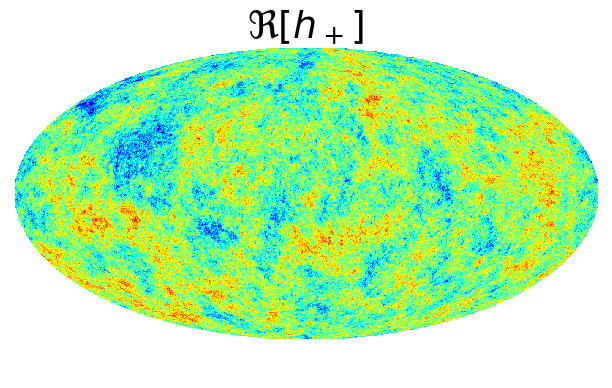}%\hfil
    \includegraphics[width=0.25\linewidth]{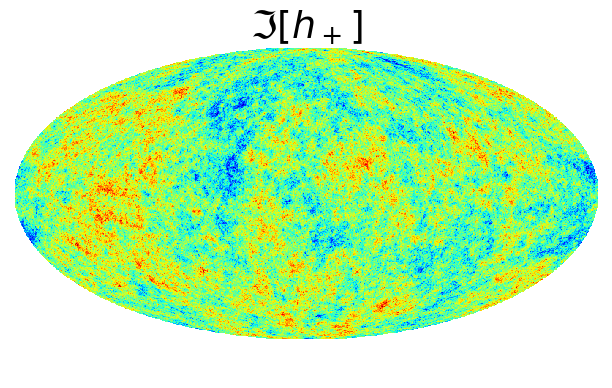}%\par\medskip
    \includegraphics[width=0.25\linewidth]{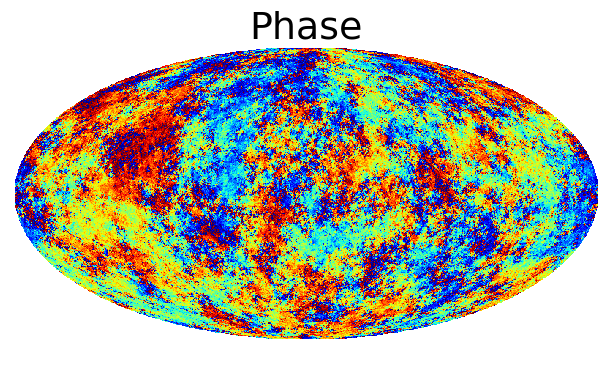}%\hfil
    \includegraphics[width=0.25\linewidth]{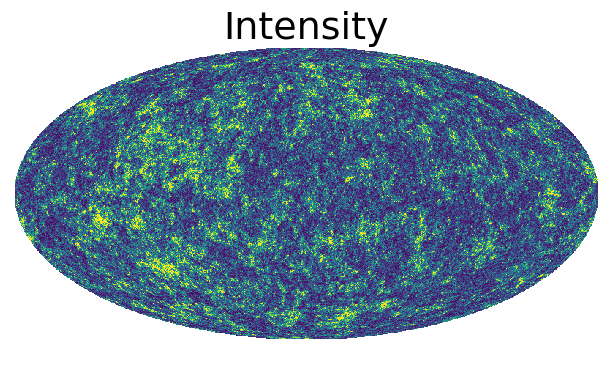}\par\medskip
    \includegraphics[width=0.25\linewidth]{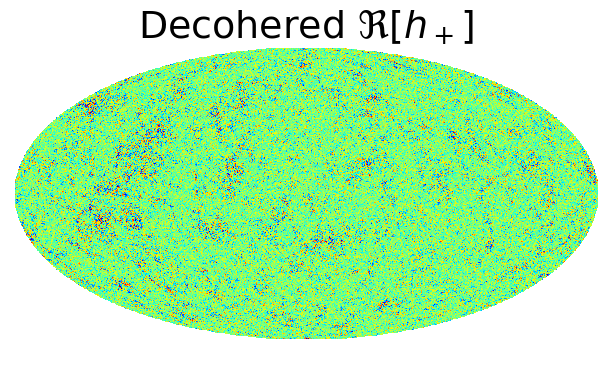}%\hfil
    \includegraphics[width=0.25\linewidth]{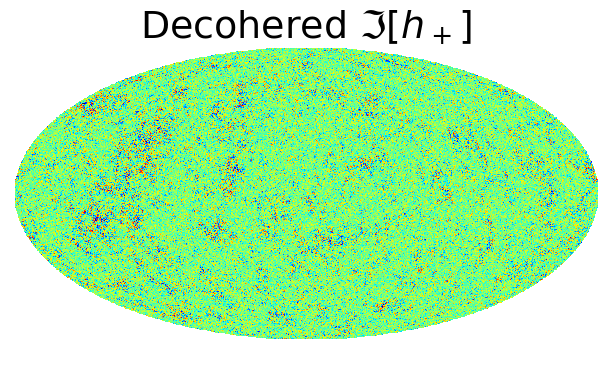}%\par\medskip
    \includegraphics[width=0.25\linewidth]{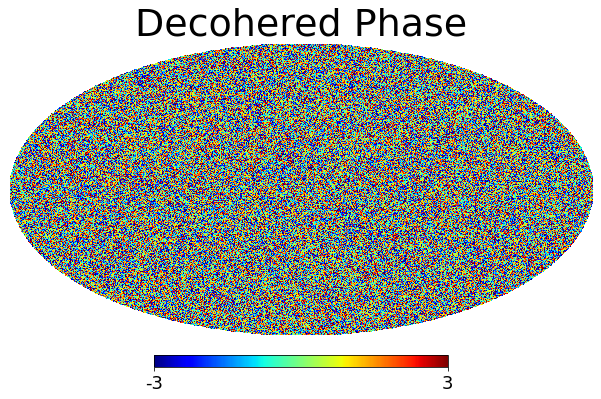}%\hfil
    \includegraphics[width=0.25\linewidth]{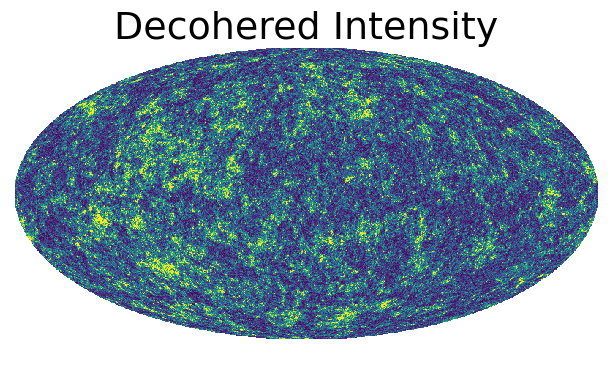}\par\medskip
    \includegraphics[width=0.25\linewidth]{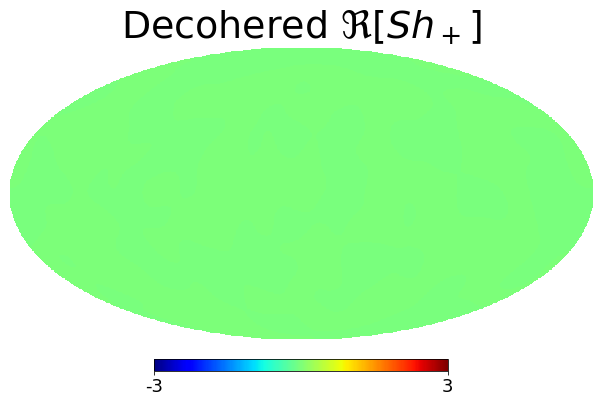}%\hfil
    \includegraphics[width=0.25\linewidth]{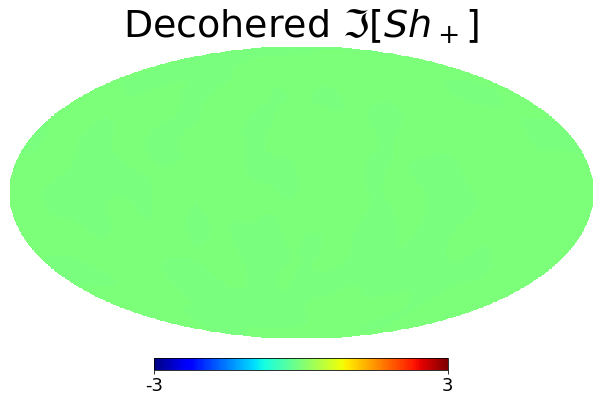}%\par\medskip
    \includegraphics[width=0.25\linewidth]{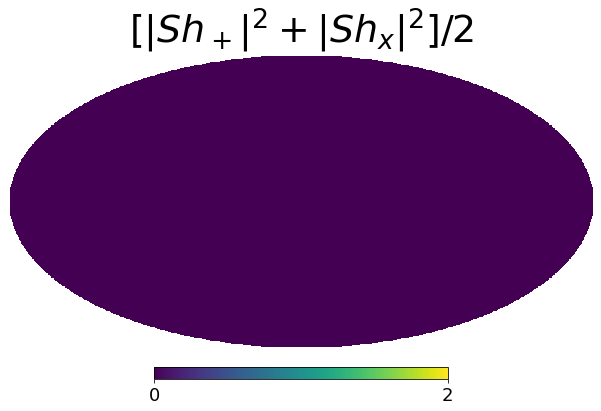}%\hfil
    \includegraphics[width=0.25\linewidth]{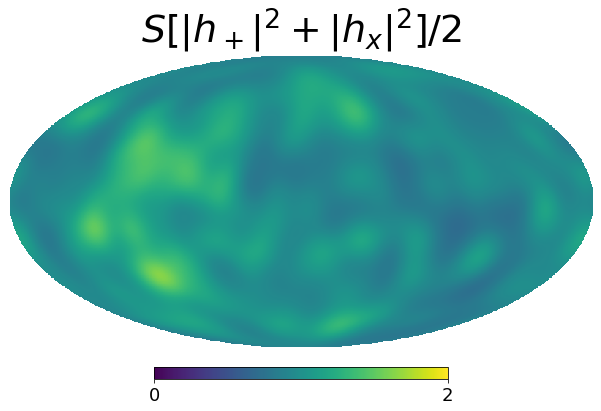}
\caption{A representation of the level of decoherence at frequencies $10^{-9}$ Hz. {\sl Top, from left to right}: A realisation of real and imaginary components of strain $h$ (we show only $h_+$) for an $\ell^2 C_\ell\sim $ constant SGWB. The maps are normalised to unit variance. The corresponding phase map is shown colour-coded in the range $-\pi$ to $\pi$ showing the angular correlations present in the original signal. The intensity $I=(|h_+|^2+|h_\times|^2)/2$ of the original background is shown on the right. {\sl Middle}: The phase of the signal is modified by the decoherence field obtained from a realisation of the spectrum shown in Fig.~\ref{fig: plot of C_l}. The phases are now randomised and the real and imaginary components of $h_+$ have been rotated randomly on across the sky. The intensity is unaffected. {\sl Bottom}: Smoothing the maps using a full width half maximum $\sim 20$ degree kernel as a proxy for observations shows how the structure in $h_+$ is erased by averaging over the random phases. The final two maps show the intensity obtained from the smoothed $h$ maps and from a smoothing of the intensity obtained from the original $h$ maps respectively. The final map is a proxy for an incoherent reconstruction method using estimators for the intensity $I$ instead of strain $h$.}\label{fig: maps}
\end{figure*}

The {\sl Weyl} potential, defined as the combination $\Phi_W = (\Phi + \Psi)/2$, determines the overall effect. The evolution of $\Phi_W$ in the Fourier domain can be computed using Einstein-Boltzmann solvers such as {\tt CAMB}\footnote{\url{https://camb.info/}} \cite{lewis1999}. This allows us to define an angular transfer function for a l.o.s. calculation of the angular power spectrum for the quantity $\delta\varphi$
\begin{equation}\label{eq: C_l power spectrum}
    C^{\delta\varphi}_\ell = 32 \pi\, f^2 \int_{0}^{\infty}\textrm{d} k\; k^2 P_W(k) |\Delta^W_\ell(k,\eta_o)|^2\,,
\end{equation}
where $(2\pi)^3\delta^{(3)}(\mathbf{k}-\mathbf{k}')P_W(k) = \langle \Phi^0_W(\mathbf{k})\Phi^{0\star}_W(\mathbf{k}')\rangle$ is the primordial power spectrum of the  potential. The angular transfer function integrated to $\eta_o$ is defined as\footnote{The transfer function is not stochastic and is independent of direction on the sky.}
\begin{equation}
    \Delta^W_\ell(k,\eta_o) = \int_{\eta_e}^{\eta_o}  \textrm{d}\eta\, j_\ell[k(\eta_o - \eta)]  \frac{\Phi_W(\eta, k)}{\Phi^0_W(k)}\,,
\end{equation}
where $j_\ell(x)$ are spherical Bessel functions.

A measure of the typical time delay is given by the one-point correlation of the phase delay
\begin{equation}\label{eq: rms time delay}
    d_{\text{rms}}^2 \equiv \langle d^2(\n)\rangle = \frac{1}{(2 \pi\, f)^2}\sum_{\ell=1}^{\infty} \frac{2\ell+1}{4 \pi} C^{\delta\varphi}_\ell\,, 
\end{equation}
where angular brackets, in this case, denote sky averaging. This angular dependence of the delay is in addition to an isotropic ($\ell=0$) contribution due to the background expansion that is zeroth order in the scalar perturbation.

A typical angular power spectrum for a best-fit $\Lambda$CDM cosmology is plotted in Fig.~\ref{fig: plot of C_l} for a range of source redshifts. 
The spectrum converges by $z_e\sim 1000$. 
At high redshifts, the resulting rms delay is $d_{\text{rms}} \sim 0.8$ Mpc, in agreement with earlier estimates \cite{hu2000} for CMB photons. 
In the best case scenario, one might hope that low frequencies would mitigate the impact of this large distance on the phase. Unfortunately, even for nanohertz frequencies, characteristic of the PTA band, the expected phase shift is $\delta \varphi \sim 10^5$. A realisation of $\delta\varphi$ for $z_e=1000$ at $f=10^{-9}$ Hz is shown in Fig.~\ref{fig:dvarphi}. This leads to a randomisation of the number of wave cycles in each direction of the sky\footnote{Information encoded in phase modulations is unrecoverable unless the linear limit $\delta\varphi\ll 2\pi$ is maintained.}. Our results suggest that the phase shift due to the cosmological Shapiro time delay could only be treated as a linear perturbation for frequencies below $10^{-12}$ Hz for any source at a redshift $z_e \gtrsim 0.01$. Since the effect is cumulative along the l.o.s., the maximum frequency for which $\delta \varphi_{\text{rms}} \lesssim 1$ decreases as the redshift to the source increases.

{\sl Decoherence on the sky.--} We now show how this result affects the information contained in the GW strain for an initially coherent background. For completeness, we note that the strain amplitude $\mathcal{A}$ also receives corrections from l.o.s. effects. Perturbations therefore introduce an angular dependence in the {\sl intensity} of the background which is linear in the scalar potentials. The information imprinted in the intensity is preserved and has already been considered elsewhere \cite{laguna2010,contaldi2017,cusin2017}.

Fig.~\ref{fig: maps} illustrates the consequence of decoherence for the $h_{+}$ component of a SGWB. The  maps for $h_+$ and $h_\times$ (not shown here) are obtained using spin-2 realisations of a constant $\ell^2C_\ell$ power spectrum of equal amplitude for both grad and curl modes \cite{Stebbins:1996wx}\footnote{\url{https://healpix.jpl.nasa.gov/}}. The real and imaginary components of $h_+$ are shown along with the phase (top row). The phase shows angular correlations due to the coherence of the original field. We create a decoherence field as a spin-0 realisation of the spectrum shown in Fig.~\ref{fig: plot of C_l} and apply it to the original $h$ maps by rotating the components using the resulting phase shift (middle row). We ignore the effect of perturbations on the amplitude. Since the typical $\delta\varphi \gg 2\pi$ in any direction the final map is effectively a random rotation of the original complex mode and any original coherence is erased, as shown in the final phase map (middle, second from right). The information carried by the angular correlations in the amplitude is preserved, however.

{\sl Mapping SGWBs.--} It is important to understand the consequence of the decoherence effect on how one should {\sl estimate} the underlying signal using detectors with finite resolution. We note here explicitly that coherent detectors (such as LIGO, Virgo, LISA etc.) output a data stream which can be interpreted using either coherent or incoherent mapping methods. Coherent methods involve estimators that are linear in strain $h$ and attempt to reconstruct $h_+$ and $h_\times$ maps together with their complex phases. However, considering the estimator is effectively an averaging (smoothing) of the strain signal across the sky, these will estimate quantities that vanish in the ensemble limit (see bottom row of Fig.~\ref{fig: maps}). An incoherent method, based on a cross correlation of the data (order $h^2$), estimates the intensity of the underlying signal and discards any phase information from the outset. The ensemble average of the intensity does not vanish. The difference is analogous to the estimate of the variance of a centred random variate. One cannot estimate the variance by taking the square of the sum of random draws of the variate. Rather, one takes the sum over the squares of individual random draws. The same problem is present when averaging over modes of different wavelength instead of direction. This will inevitably be the case for finite duration measurements which limit the frequency resolution.

Since the decoherence is effectively {\sl complete}, even for originally coherent SGWBs, we argue there is no use in building coherent estimation methods for mapping diffuse SGWBs. Only incoherent methods are of any use in mining the information contained in anisotropies of SGWBs through the intensity of the background. As we have noted, intensity is conserved to leading order in anisotropies, and thus remains a bountiful source of information about GWs at emission. This is shown in Fig.~\ref{fig: maps} for the case where the ``estimation'' (smoothing) step is applied only {\sl after} the phase information is discarded (bottom right).

This result has practical consequences with respect to map-making algorithms. In particular, for LISA, where noise is cross-correlated between channels, estimators must be based on the maximisation of a full likelihood in the data cross-correlation. This will require a numerical search for the maximum likelihood solution since there is no $\chi^2$-derived, closed-form solution.

{\sl Discussion.--} We have shown that the ability of phase-coherent mapping methods to reconstruct images of GW sources is fundamentally limited by the small wavelength of the observed modes as compared to cosmological distances. In a homogeneous universe, phase is a conserved quantity along the null geodesic of a GW. 
If we lived in such a universe, phase-coherent mapping \cite{cornish2014, gair2014, romano2015} could produce a faithful depiction of the GW sky.
Allowing for metric perturbations, however, leads to a large, non-linear rotation of the complex components of the GW modes. Phase information is therefore randomised such that no angular correlation remains in a linear signal for frequencies above $10^{-12}$ Hz. 

On the other hand, l.o.s. corrections to the intensity and spectrum \cite{laguna2010,contaldi2017,domcke2020} can be treated as first order perturbations, making these quantities a valuable source of information about both GW creation mechanisms and the gravitational potentials which perturb them.

The results obtained in this article are not only relevant for mapping techniques, but also for efforts to characterise SGWBs with higher-order statistics~\cite{Bartolo:2018evs,Bartolo:2018qqn,Tsuneto:2018tif, Powell:2019kid}. 
Correlators, and hence estimators, built from an odd number of $h$'s are expected to be zero due to residual phase dependence causing the sky-average to disappear.
Then, as per \cite{bartolo2019a, bartolo2019c, Bartolo:2019zvb}, it is only meaningful to define observables by correlating quantities which are quadratic in $h$ (and thus phase-independent).

{\sl Acknowledgements.--}
We thank Giulia Cusin, Valerie Domcke, Marco Peloso and Angelo Ricciardone for useful discussions. We also acknowledge Arianna Renzini for detailed discussions about incoherent map-making.
This research was supported by Science and Technology Facilities Council consolidated grant ST/P000762/1. AM is funded by the President's Ph.D. Scholarships. 

\bibliographystyle{apsrev}
\bibliography{refs.bib}
\end{document}